\documentclass{article}
\usepackage{spconf,amsmath,graphicx}
\usepackage{booktabs}

\title{On real-time multi-stage speech enhancement systems}
\name{
    \parbox{\linewidth}{\centering
   Lingjun Meng$^{1,2}$ Jozef Coldenhoff$^{1,2}$ Paul Kendrick$^{2}$ Tijana Stojkovic$^{2}$ Andrew Harper$^{2}$\\
     \textit{Kiril Ratmanski$^{2}$, Milos Cernak$^{2*}$}
    \thanks{$^*$Corresponding Author: milos.cernak@ieee.org}
    }
}

\address{$^{1}$Ecole Polytechnique Federale de Lausanne(EPFL), Lausanne, Switzerland\\
$^{2}$Logitech Europe S.A., Lausanne, Switzerland\\
}

\begin{document}
\ninept

\maketitle

\begin{abstract}
Recently, multi-stage systems have stood out among deep learning-based speech enhancement methods. However, these systems are always high in complexity, requiring millions of parameters and powerful computational resources, which limits their application for real-time processing in low-power devices. Besides, the contribution of various influencing factors to the success of multi-stage systems remains unclear, which presents challenges to reduce the size of these systems. In this paper, we extensively investigate a lightweight two-stage network with only 560k total parameters. It consists of a Mel-scale magnitude masking model in the first stage and a complex spectrum mapping model in the second stage. We first provide a consolidated view of the roles of gain power factor, post-filter, and training labels for the Mel-scale masking model. Then, we explore several training schemes for the two-stage network and provide some insights into the superiority of the two-stage network. We show that the proposed two-stage network trained by an optimal scheme achieves a performance similar to a four times larger open source model DeepFilterNet2 \cite{schroter2022deepfilternet2}. 
\end{abstract}

\section{Introduction}
\label{sec:intro}

Speech enhancement, intended to improve the quality of speech signals corrupted by additive noise, is a fundamental task in audio signal processing with various applications, including mobile telecommunication, robust automatic speech and speaker recognition \cite{wang2018supervised}, and hearing aids. Over the past decade, speech enhancement approaches based on deep learning have seen large attention and success. The reported deep speech enhancement models either predict the magnitude-only target such as magnitude spectrum and ideal ratio mask \cite{wang2014training, valin2020perceptually, valin2018hybrid}, or phase-related targets like complex ratio mask \cite{hu2020dccrn, williamson2015complex}, phase-sensitive mask \cite{erdogan2015phase}, and complex spectrum \cite{tan2019complex}. Recently, a series of multi-stage systems dominates the mainstream benchmarks \cite{ju2022tea, li2021icassp, ju2023tea, yan2023npu}. They usually predict magnitude-only targets in the first stage and then predict complex spectrum in the second stage. However, these reported multi-stage networks normally have an overall size larger than 2M, which hinders their deployment on battery-operated devices like earbuds and headsets. Besides, the contributions of individual components to the overall performance of the multi-stage systems remain unclear, which poses challenges for resizing the multi-stage networks. Focusing on real-time and on-device applications, we provide a consolidated view of the roles of each individual component of multi-stage systems based on a lightweight two-stage network with 560k total parameters. 

The Mel-scale is a perceptual scale imitating the frequency resolution of the human hearing system, which intendeds to tackle the perceptual nonlinearity in the linear frequency scale \cite{kang2018dnn}. It compresses the redundant high-frequency information and amplifies the critical low-frequency information. By using the Mel-scale spectrum, we can use fewer frequency bands without losing much important frequency information, and thus significantly reduce the model size while not degrading the performance a lot compared with the linear scale model. Based on this idea, we first extensively investigated a masking model that predicts the oracle gains \cite{valin2020perceptually, valin2018hybrid} in Mel scale. After magnitude masking, gain post-filters are frequently used \cite{schroter2022deepfilternet2, valin2020perceptually, valin2018hybrid}. However, sometimes the gain post-filters instead cause performance degradation. By understanding the roles of post-filters and the gain power factor in the training loss, we found that the post-filter only contributes positively when it performs a complementary role to the gain power factor. Besides, inspired by \cite{pandey2018new}, we include the post-processing steps of Mel-scale masking into the training graph and supervise the model training with time-domain loss functions, which significantly reduces the speech distortion and further enhances the perceptual speech quality, especially in low signal-to-noise ratio (SNR) scenarios.

Nonetheless, there exist two sources of distortion inherent in Mel-scale masking. The first is a frequency information loss caused by Mel compression, while the second is phase-related loss due to using a noisy phase when reconstructing the enhanced speech. The former will be shown to be insignificant compared to (uncompressed) frequency-scale masking. However, the phase-related distortion will be demonstrated as significant, especially in low SNR scenarios. Thus, a second-stage model, which predicts the clean complex spectrum from the first-stage enhanced spectrum and the original noisy spectrum, is introduced to reduce the phase-related distortions and further suppress the residual noise. After a consolidated exploration of two-stage training schemes, we obtained a two-stage model that significantly improved the enhanced speech quality and reduced speech distortions beyond the single-stage models. Moreover, the proposed two-stage model performs similarly to a four times larger open-source two-stage model DeepFilterNet2~\cite{schroter2022deepfilternet2}. 
\begin{figure*}[htb]

  \centering
  \centerline{\includegraphics[width=1.75\columnwidth]{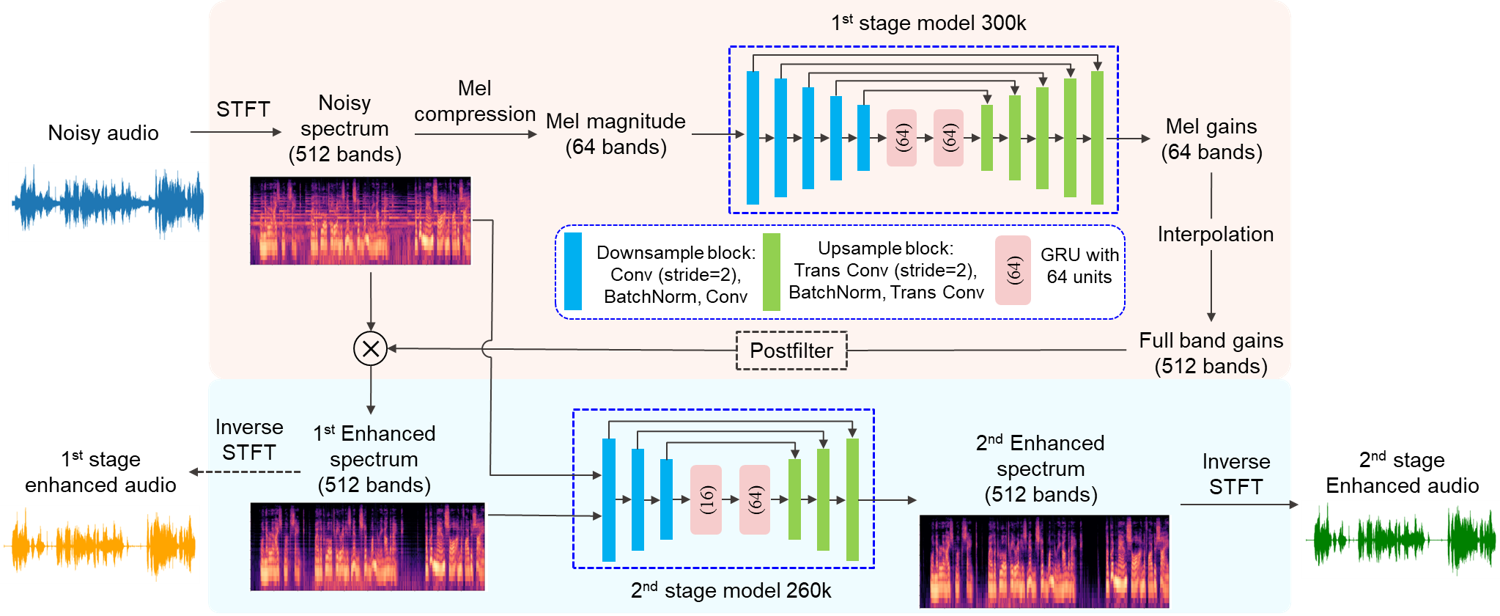}}
\caption{Two-stage speech enhancement pipeline. Top: masking-based magnitude stage, bottom: mapping-based complex stage.}
\label{pipeline}
\end{figure*}

\section{Light-weight Two stage speech enhancement system}
\label{sec:methods}

\subsection{Speech enhancement pipeline}
The two-stage speech enhancement pipeline is shown in Figure \ref{pipeline}. After the Short-Time Fourier Transform (STFT) of the noisy signal, the derived magnitude spectrum in linear frequency scale is converted to Mel scale by applying Mel filter bank. The Mel spectrum is compressed through a natural logarithm before being fed into the neural network. The first stage, the Mel-scale masking model, is trained to predict the Mel gains.
The oracle Mel gains are defined by Eq.~\eqref{oracle_mel_eq}, where $|S|$ denotes the target speech magnitude, $|X|$ denotes the noisy mixture magnitude, and the subscript $Mel$ denotes corresponding quantities in the Mel scale.
\begin{equation}
\label{oracle_mel_eq}
    g_{Mel}=\frac{|S|_{Mel}}{|X|_{Mel}}
\end{equation}

The predicted Mel gains must be interpolated back to the linear frequency scale before being applied to the noisy spectrum. The interpolation is done through the transpose of the Mel filter bank. The mask can be applied directly or after post-filtering. After masking the noisy magnitude, the noisy phase is combined with enhanced magnitude to obtain the enhanced spectrum and then to obtain the first stage enhanced speech audio.

After the first stage network, the enhanced complex spectrum is concatenated with the original, noisy complex spectrum and then fed into the second stage network to predict the clean complex spectrum. We separate the complex spectrum into the real and imaginary parts and then concatenate the real and imaginary parts for input and output of the network. Thus, the second stage network is still real-valued. 

\subsection{Model architecture}
\label{arch}

As audio signals are time-dependent objects, recurrent networks would be natural options for deep speech enhancement models \cite{wang2018supervised, pandey23b_interspeech}. However, Sach et al. \cite{sach23_interspeech} demonstrated that the convolutional recurrent networks show a significant advantage compared with recurrent networks. Tan et al. \cite{tan2018convolutional} reported an efficient architecture that combines Unet convolution layers with long short-term memory networks (LSTM). Inspired by this work, we designed an Unet-GRU architecture, replacing the heavy LSTM with more efficient Gated Recurrent Units (GRU). Focusing on real-time applications, causal convolution is used in both networks. The kernel size is 1 (the time axis) and 3 (the frequency axis). Skip connections of Unet are designed to avoid the gradient vanishing issues. 

As shown in Figure \ref{pipeline}, the Unet-GRU architecture is used in both stages. The first stage network has 300k trainable parameters. There are five downsample blocks and five corresponding upsample blocks. Each downsample block contains two convolution layers and 1 batch normalization layer. Each upsample block contains two transposed convolution layers and 1 batch normalization layer. The input is single-channel with 64 Mel bands, and the initial channel after the first downsample block is 8. There are two recurrent layers between the encoder and decoder, with 64 GRUs in each layer. The second stage model takes the input of 512 linear bands (full band) with four channels and has 260k parameters. It has 16 initial channels after the first downsample block, three downsample blocks, and three upsample blocks. There are 16 GRUs in the first recurrent layer and 64 GRUs in the second recurrent layer. 

\subsection{Loss functions and gain post-filter}

For the magnitude masking models, the gain loss function is commonly used \cite{valin2020perceptually, valin2018hybrid}. It is defined by \eqref{gain_loss}, which is the mean square error of oracle gains $g$ and estimated gains $\hat{g}$ to the power of $p$. As gain values range in $[0,1]$, the power factor $p$ controls which bins should be emphasized during learning. For example, if $p=0.5$, the contribution of the relatively noisy bins where $g$ is close to $0$ is amplified, while the contribution of relatively clean bins where $g$ is close to $1$ is reduced. In this case, the model is encouraged to attenuate more noise. On the other hand, if $p=2$, the model is trained to mildly attenuate noise while paying more attention to remaining speech.
\begin{equation}
\label{gain_loss}
    L_g=\frac{1}{TF}\sum_t^T\sum_f^F (g^{p}_{t,f}-\hat{g}^{p}_{t,f})^2
\end{equation}

After predicting the Mel gains and interpolating back to the frequency scale, an element-wise post-filter \cite{valin2020perceptually, schroter2022deepfilternet2}, defined in Eq.~\eqref{sin_pf}, is sometimes used before applying gains on the noisy magnitude. The post-filter is designed to attenuate noise in the noisy frequency bins further while maintaining the predicted gain values in the other bins. Thus, the post-filter has a similar role of the gain loss function with $p=0.5$ and contributes complementarily to the gain loss with $p=2$.

\begin{equation}
\label{sin_pf}
    g'=\hat{g}\sin{\left(\frac{\pi}{2}\hat{g}\right)}
\end{equation}



\begin{figure}[htb]
\begin{minipage}[b]{.43\linewidth}
  \centering
  \centerline{\includegraphics[width=3.7cm]{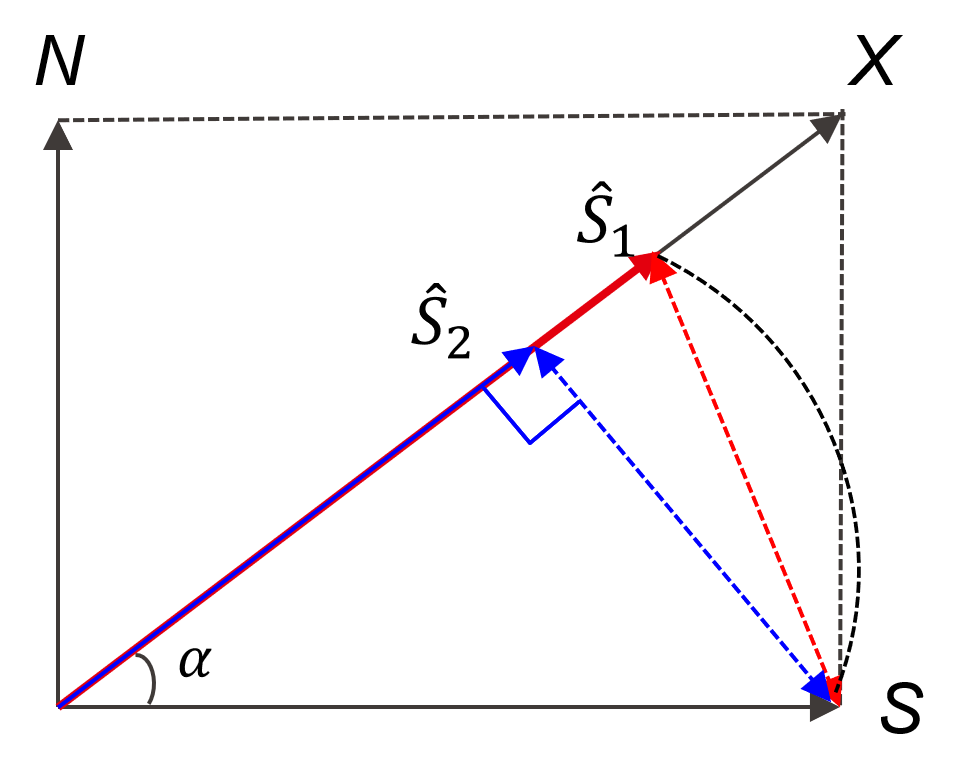}}
  \centerline{(a) Spectral}\medskip
\end{minipage}
\hfill
\begin{minipage}[b]{.53\linewidth}
  \centering
  \centerline{\includegraphics[width=4.7cm]{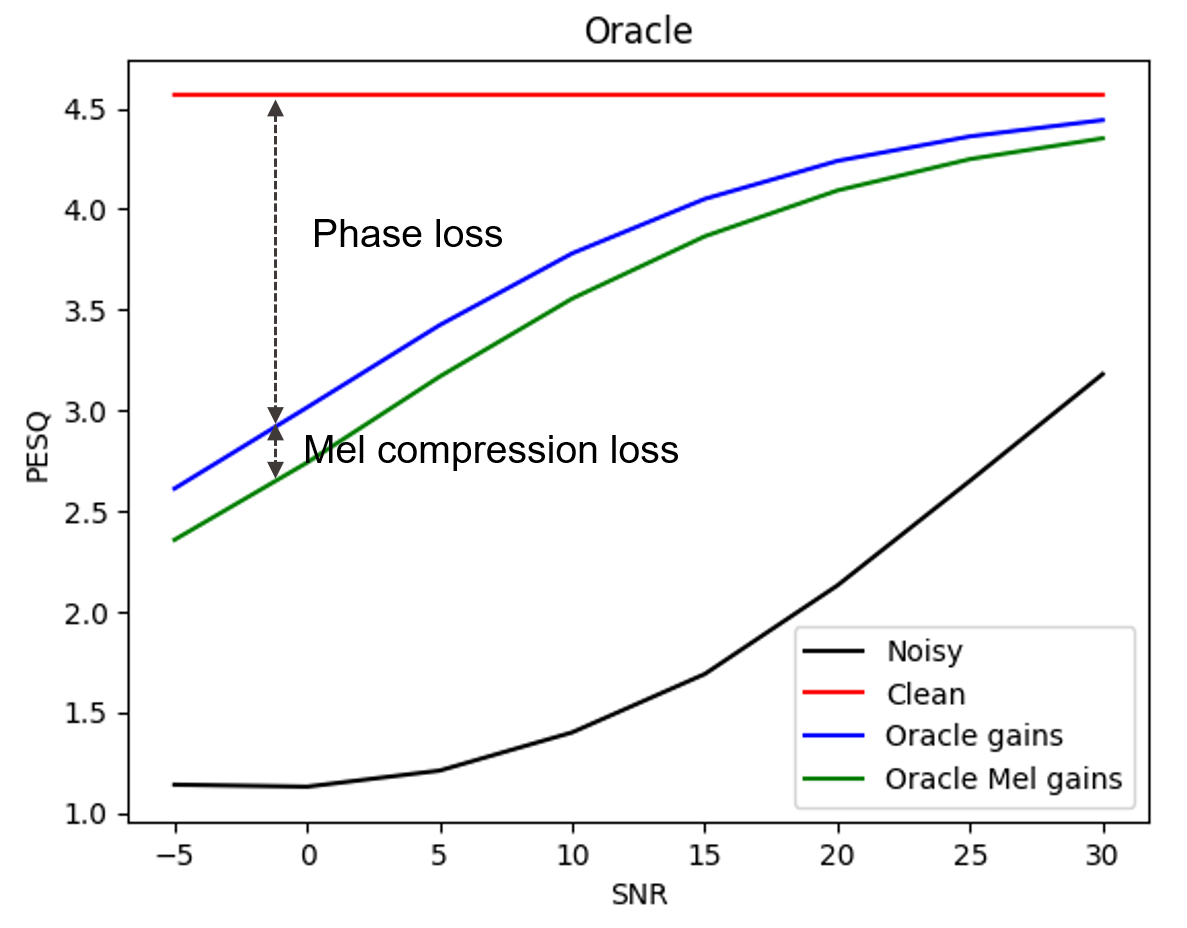}}
  \centerline{(b) Qualitative}\medskip
\end{minipage}
\caption{Illustration of phase importance: a) Spectral ($N$: additive noise; $S$: target speech; $X$: noisy mixture; $\hat{S}_1$: ideal spectral vector estimated by using magnitude-only loss; $\hat{S}_2$: spectral vector using noisy phase and closest to target speech), b) Qualitative (phase-related loss obtained with oracle inference).}
\label{fig:phase}
\end{figure}

In addition to the post-filter Eq.~\eqref{sin_pf}, other DSP-based post-filter approaches can be applied in the first stage network. As part of this work, we have also applied a spectral weighting method described in \cite{Gustafsson1998ANP}. By exploiting the masking properties of the human auditory system, the spectral weighting rule is designed to mask the distortions of the residual noise. In our case, however, this post-filter has not given better performance compared to the ``sin'' post-filter.

Supervising the Mel-scale masking model by gain loss will ignore all the phase-related information, which could lead the model to a suboptimal point with respect to the distance to the target speech, as shown in Figure \ref{fig:phase}a. The bias could be evident when the angle $\alpha$ is large, i.e., in low SNR scenarios. In fact, the differentiability of the interpolation and inverse STFT makes it possible to supervise the Mel-scale masking model by time-domain losses like SI-SNR defined in Eq.~\eqref{sisnr}, where $s$ and $\hat{s}$ are the target and estimated audio signals, respectively. Beyond $L_{SI-SNR}$, magnitude loss $L_{mag}$ and asymmetric loss $L_{asym}$ are also added following \cite{ju2022tea}. Note that the $\beta$ in $L_{mag}$ and $L_{asym}$ is the power compression factor set to 0.5 in this work. It is intended to compress the range of magnitude so that the model is easier to resolve the low power noise, which is different from the gain power factor $p$.

\begin{equation}
    L_{mag}=\frac{1}{TF}\sum_t^T\sum_f^F (|S|^{\beta}_{t,f}-\hat{|S|}_{t,f}^{\beta})^2
\end{equation}

\begin{equation}
    h(x) = 
     \begin{cases}
       0 &\quad\text{if}\ x\leq 0\\
       x &\quad\text{if}\ x > 0\\
     \end{cases}
\end{equation}
\begin{equation}
\label{asym}
    L_{asym}=\frac{1}{TF}\sum_t^T\sum_f^F \left|h(|S|_{t,f}^{\beta}-\hat{|S|}_{t,f}^{\beta})\right|^2
\end{equation}

\begin{equation}
\label{sisnr}
    L_{\text{SI-SNR}}=-10\log_{10}\left(\frac{\|\kappa s\|^2}{\|\kappa s-\hat{s}\|^2}\right)
\end{equation}
\begin{equation}
\label{scale}
    \kappa=\arg\min_{\kappa}\|\kappa s-\hat{s}\|^2=\frac{\hat{s}^Ts}{\|s\|^2}
\end{equation}
\begin{equation}
\label{snr_total_loss}
    L_1=(L_{mag}+L_{asym})\cdot F+2L_{\text{SI-SNR}}
\end{equation}

There are also many loss options for complex spectrum learning. Sebastian et al. \cite{braun2021consolidated} made a consolidated view for the commonly used loss items and concluded that combining magnitude-only with phase-aware objectives always leads to improvements. Thus, we also combine magnitude loss and phase-aware loss as our second-stage loss $L_2$, which is also the final loss of our two-stage network. The phase-aware loss is defined in Eq.~\eqref{eq_phase}, where $\theta_{t,f}$ and $\hat{\theta}_{t,f}$ are the phase of target and estimated spectrum in each bin.

\begin{equation}
\label{eq_phase}
    L_{phase}=\frac{1}{TF}\sum_t^T\sum_f^F (|S|^{p}_{t,f}\exp(i\theta_{t,f})-\hat{|S|}_{t,f}^p\exp(i\hat{\theta}_{t,f}))^2
\end{equation}
\begin{equation}
    L_{2}=L_{mag}+L_{phase}
\end{equation}

\section{Experiment results and analysis}
\label{result}

\subsection{Experiment setup}

The training samples and test samples are generated from the DNS Challenge dataset by randomly mixing the speech samples and noise samples. We have $72000$ samples for training and $500$ samples for validation during training. Each audio is in 10 seconds. Thus, the total length of training samples is 200 hours. The sample rate is 32 kHz, and the hop length is 320. Thus, there are 1000 time frames in the spectrum. We used the standard Hann window with a length of 640. For test samples, as we would like to observe the performance under different SNR scenarios, we generate a noisy mixture with specific SNR. We have 100 independent mixtures of speech and noise, and each mixture has 8 SNR bands ranging from -5 to 30 with a step of 5. Thus, we have 800 test samples in total. The enhanced speech performance is evaluated by PESQ \cite{rix2001perceptual} and SI-SDR \cite{le2019sdr}.
\subsection{Mel-scale masking model investigation}
In order to understand the systematic distortions of the Mel-scale masking model, we first conducted oracle inference, i.e., the performance enhanced by oracle gains. The oracle inference results are shown in Figure \ref{fig:phase}b. The gap between oracle gains and oracle Mel gains is the loss caused by compressing 512 linear bands to 64 Mel bands, which is much smaller than the loss caused by ignoring phase information. The phase-related loss is especially significant in low SNR scenarios. The oracle inference results partially demonstrated the rationality of Mel compression, which remarkably reduces the model size while causing relatively small systematic distortions. The oracle inference results confirm the importance of phase information, particularly in low SNR scenarios.

\begin{figure}[htb]
\begin{minipage}[b]{\linewidth}
  \centering
  \centerline{\includegraphics[width=7.0cm]{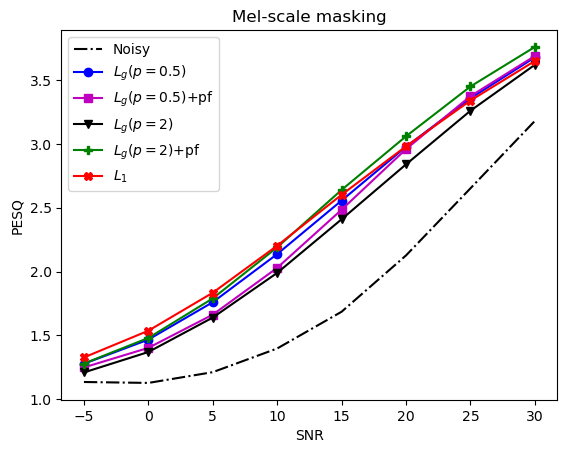}}
  \centerline{(a) Speech quality}\medskip
\end{minipage}

\hfill
\begin{minipage}[b]{\linewidth}
  \centering
  \centerline{\includegraphics[width=7.0cm]{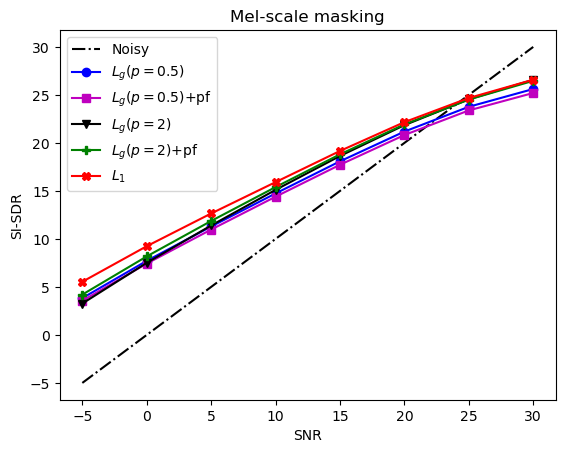}}
  \centerline{(b) Speech distortion ratio}\medskip
\end{minipage}
\caption{The impact of the gain power factor $p$, the postfilter $pf$ and loss functions on: a) PESQ, b) SI-SDR}
\label{mel_infer}
\end{figure}

The inference results for the Mel-scale masking model are shown in Figure \ref{mel_infer}. As discussed before, when the gain power factor $p$ is set to 0.5, the model pays more attention to noisy bins and tends to attenuate the noise aggressively. In this case, applying the post-filter could degrade the enhanced speech quality and increase speech distortion due to their overlapping roles. On the other hand, as the model trained by gain power factor $p=2$ pays more attention to cleaner bins, the noise is not sufficiently canceled. Thus, the enhanced speech quality is significantly lower than the model trained by $p=0.5$ in terms of PESQ. However, after applying the post-filter to attenuate more noise in noisy bins based on the prior knowledge from the model, the PESQ is significantly enhanced. More interestingly, due to the complementarity, the postfiltered gains perform even better than the aggressive denoising model trained by $p=0.5$. As the model trained by $L_1$ can see the phase information, the speech distortions are evidently reduced according to SI-SDR results, and PESQ is also improved in low SNR compared with the one learned by gain loss. The models could degrade the SI-SDR in very high SNR as SI-SDR is not sensitive to residual noise and mainly measures speech distortions.

\subsection{Two-stage training schemes comparison}

The inference results for different two-stage training schemes are listed in Table \ref{two-stage-scheme}. Direct joint training schemes perform moderately because the first stage model could not be optimized sufficiently due to the gradient vanishing issues. The separate training scheme with the first stage model trained by gain loss performs even worse than direct joint training and significantly worse than other schemes. In contrast, the separate training scheme with the first stage model trained by $L_1$ performs similar to the optimal one. These indicate that the first stage model trained by $L_g$ still has a distance to the two stage global optimum while the one trained by $L_1$ almost reaches the global optimum. Other training schemes performs similarly, which indicates that these two-stage models almost reach the optimum. By comparing both PESQ and SI-SDR, the best training scheme is considered to be training the first-stage model with $L_1$ and then jointly training the two-stage model and the trained two-stage model is used for the following comparison.
\begin{table}[htb]
    \centering
    \caption{Two stage training schemes comparison ("$\pm$" represents 95\% confidence intervals)}
    \begin{tabular}{lcc}
    \toprule
        Training schemes & PESQ & SI-SDR  \\
        \midrule
        joint & 2.53 $\pm$ 0.05 & 17.15 $\pm$ 0.55  \\ 
        stage1($L_g$)-stage2 & 2.52 $\pm$ 0.07 & 16.89 $\pm$ 0.59  \\ 
        stage1($L_g$)-joint & 2.57 $\pm$ 0.06 & 17.51 $\pm$ 0.56\\ 
        stage1($L_g$)-stage2-joint & 2.59 $\pm$ 0.07 & 17.44 $\pm$ 0.58 \\
        stage1($L_1$)-stage2 & 2.57 $\pm$ 0.05 & 17.57 $\pm$ 0.56  \\ 
        stage1($L_1$)-joint & 2.58 $\pm$ 0.06 & 17.58 $\pm$ 0.55  \\ 
        stage1($L_1$)-stage2-joint & 2.59 $\pm$ 0.07 & 17.51 $\pm$ 0.58  \\
    \bottomrule
    \end{tabular}
    \label{two-stage-scheme}
\end{table}

To demonstrate the superiority of our proposed two-stage network, we trained a single-stage Unet-GRU complex spectrum mapping model, and then we tested the model as well as a baseline model DeepFilterNet2 \cite{schroter2022deepfilternet2}
on our test dataset. The DeepFilterNet2 is an open source model also based on two-stage learning. It was originally proposed for real-time speech enhancement but also applied in hearing aids \cite{schroter23_interspeech}. The state-of-the-art single stage complex model was proposed by Ristea et al.\cite{ristea23_interspeech}, but the size exceeds 7.5M. Hence we still use our Unet-GRU based single stage complex model for comparison. The overall results with statistical significance test are listed in Table \ref{summary}. We found that the two-stage model significantly outperforms both single-stage models in terms of PESQ and SI-SDR. Besides, our two-stage model performs similarly to DeepFilterNet2, which has 2M parameters. The complex spectrum without Mel transformation contains much redundant information, which hinders sufficient learning under limited model capacity. These results potentially demonstrate the superiority of a two-stage framework with Mel-scale masking in the first stage and complex spectrum learning in the second stage.
\begin{table}[!ht]

    \centering
    \caption{Comparison of two-stage model, single-stage models, and baseline model ("$\pm$" represents 95\% confidence intervals)}
    \begin{tabular}{lccc}
    
    \toprule
             & PESQ & SI-SDR & \#Params\\ \midrule
        noisy & 1.81 $\pm$ 0.05 & 12.50 $\pm$ 0.82 & -\\ 
        DeepFilter2 & \textbf{2.60 $\pm$ 0.05} & \textbf{16.99 $\pm$ 0.50} & 2.1M\\
        \midrule
        Mel($L_1$) & 2.43 $\pm$ 0.06 & 16.99 $\pm$ 0.53 & 300k\\ 
        Complex & 2.45 $\pm$ 0.07 & 17.01 $\pm$ 0.54 & 600k\\
        Two-stage & \textbf{2.58 $\pm$ 0.06} & \textbf{17.58 $\pm$ 0.55} & 560k\\ 

        \bottomrule
    \end{tabular}
    \label{summary}
\end{table}

\section{Conclusions}
\label{conclusions}

This paper provides a consolidated view of the roles of individual components in a multi-stage speech enhancement system. Focusing on real-time and on-device applications, we propose a lightweight two-stage network. Through the study on the Mel-scale masking model, we demonstrate the positive contribution that could be made by complementarily combining over-attenuation post-filter and gain power factor. We also show that supervising the Mel-scale masking model with a phase-visible objective could improve speech quality and significantly reduce speech distortions compared with a magnitude-only objective due to the significant systematic phase loss in low SNR scenarios. Besides, exploring two-stage training schemes also confirms the global optimality of the phase-visible objective for first-stage model training. We also show that training the first stage model and then conducting joint training could be a potential best two-stage training scheme. Finally, we demonstrate that our two-stage model outperforms single-stage models and performs similarly to a baseline open source model DeepFilterNet2, with a four times larger size, which shows the superiority of the proposed two-stage speech enhancement framework.

\vfill\pagebreak

\bibliographystyle{IEEEbib}
\bibliography{strings,refs}

\end{document}